\documentclass[12pt]{article}

\usepackage{amsmath,amssymb,graphicx,color,array,cite}
\evensidemargin \oddsidemargin

\begin{document}

\newcommand{\story}{\vspace{5mm} \noindent $\spadesuit$ }

%%%%%%%%%%%%%%%%%%%%%%%%%%% TITLE PAGE %%%%%%%%%%%%%%%%%%%%%%%%%%%%%%
\begin{titlepage}

%-------------------- footnote symbol in title page -----------------
\renewcommand{\thefootnote}{\fnsymbol{footnote}}

%----------------------- preprint number & date ---------------------

\begin{flushright}
CQUeST-2007-0093
\end{flushright}

%---------------------------- Title ---------------------------------
\vspace{15mm}
\baselineskip 9mm
\begin{center}
  {\Large \bf Anomaly Analysis of Hawking Radiation \\
  from Acoustic Black Hole}
\end{center}

%--------------------- Authors and Addresses ------------------------
\baselineskip 6mm
\vspace{10mm}
\begin{center}
  Wontae Kim\footnote{\tt wtkim@sogang.ac.kr}
  \\[3mm]
  {\sl Department of Physics and Center for Quantum Spacetime \\
    Sogang University, C.P.O. Box 1142, Seoul 100-611, South Korea} 
  \\[10mm]
  Hyeonjoon Shin\footnote{\tt hshin@sogang.ac.kr} 
  \\[3mm] 
  {\sl Center for Quantum Spacetime, Sogang University, Seoul
    121-742, South Korea }
  \\[3mm]
\end{center}

\thispagestyle{empty}

%-------------------------- abstract --------------------------------
\vfill
\begin{center}
{\bf Abstract}
\end{center}
\noindent
The Hawking radiation from the three dimensional rotating acoustic 
black hole is considered from the viewpoint of anomaly cancellation
method initiated by Robinson and Wilczek. Quantum field near the
horizon is effectively described by two dimensional charged field
with a charge identified as the angular momentum $m$. The fluxes of 
charge and energy are obtained from the anomaly cancellation
condition and regularity at the horizon, and are shown to match with
those of the two dimensional black body radiation at the
Hawking temperature.
\\ [5mm]

\vspace{5mm}
\end{titlepage}

%%%%%%%%%%%%%%%%%%%%%%%%%%% BODY OF PAPER %%%%%%%%%%%%%%%%%%%%%%%%%%%
\baselineskip 6.6mm
\renewcommand{\thefootnote}{\arabic{footnote}}
\setcounter{footnote}{0}

In a space-time background with an event horizon, the
quantum effect of fields leads to the radiation known as
Hawking radiation \cite{hawking}.  Since Hawking radiation is a
key phenomenon in the study of black hole physics and in formulating
the theory of quantum gravity, it is always worthwhile to have
various interpretations for the purpose of deepening the theoretical
understanding of it. Recently, as an attempt to understand the nature 
of Hawking radiation, Robinson and Wilczek \cite{rw} proposed a new 
interpretation that the Hawking radiation plays the role of preserving
general covariance at the quantum level by canceling the 
diffeomorphism anomaly at the event horizon.  While there has
been a similar work \cite{cf} which is specialized to 
two-dimensional space-time, their proposal is supposed to be 
valid in any space-time dimension.

The method of anomaly cancellation \cite{rw} has been developed
by considering the static and spherically symmetric black hole.
Its extension to charged and rotating black holes has been done
in \cite{iso1,iso2}, where the original idea by Robinson and 
Wilczek has been elaborated.  The result of this elaboration is
that Hawking radiation is capable of canceling anomalies of local 
symmetries at the event horizon.  In subsequent works
\cite{murata}-\cite{wp},\footnote{For a status review, see \cite{das}}
the method of anomaly cancellation have been applied to various
kinds of black objects in various dimensions including black ring.
Up to now, all the results have obtained correct Hawking fluxes and
verified the validity of the method. We note that, as a further 
elaboration on the method itself, Hawking fluxes of higher-spin 
currents have been also derived in \cite{iso4}.

In our previous work \cite{sk}, we applied the anomaly cancellation
method to a typical black hole in string theory, the non-extremal 
D1-D5 black hole \cite{hms}, and confirmed the idea of Robinson and
Wilczek.  In this work, we consider a kind of acoustic black hole, 
the `draining bathtub' fluid flow referred to as an acoustic analogue 
of (2+1)-dimensional rotating black hole \cite{visser}.  The basic idea 
of acoustic black hole was developed by Unruh \cite{unruh} in an 
attempt to connect black hole physics to the theory of supersonic 
acoustic flows.  The acoustic analogy is usually used in the 
investigation of models proposed for the micro-physical origin of 
Hawking radiation.  This naturally leads us to ask whether the
idea of anomaly cancellation is valid even for the acoustic black
hole.  We note that the validity check also gives a chance to
put the connection between the black hole physics and the theory
of acoustic flows on a firmer footing.

The acoustic line element of the draining bathtub\footnote{For
a recent study on the acoustic black hole, see for example
\cite{kko}.  We follow the notation and convention of it}
modeled by a
(2+1)-dimensional flow with a sink at the origin is
\begin{align}
ds^2 = - c^2 dt^2 + \left( dr + \frac{A}{r} dt \right)^2
+ \left( r d\phi - \frac{B}{r} dt \right)^2 ~,
\label{metric}
\end{align}
where $c$ is the speed of sound wave, and $A$ and $B$ are arbitrary
real positive constants.  From this metric, the radial positions of
the acoustic even horizon, $r_H$, and the ergosphere, $r_e$, are 
obtained as
\begin{align}
r_H = \frac{A}{c} ~, \quad r_e = \frac{\sqrt{A^2+B^2}}{c}~.
\end{align}
In dealing with the metric, it is convenient to take a coordinate
transformation in the exterior region of $r_H < r < \infty$
\cite{basak} given by
$dt \rightarrow dt + \frac{Ar}{r^2c^2-A^2}dr$, 
$d\phi \rightarrow d\phi + \frac{AB}{r(r^2c^2-A^2)}dr$. The metric
(\ref{metric}) is then rewritten in conventional form as
\begin{align}
ds^2 = - N^2 dt^2 + \frac{1}{N^2} dr^2 + 
r^2 ( d\phi - \Omega_0 dt)^2 ~,
\label{abh}
\end{align}
where the time coordinate has been rescaled by $c$ for simplicity, and
\begin{gather}
N^2 (r) = 1 - \frac{A^2}{c^2 r^2} =1-\frac{r_H^2}{r^2}~, \notag \\
\Omega_0 (r) = \frac{B}{c r^2} = \Omega_H \frac{r_H^2}{r^2}~, \quad
\Omega_H = \frac{B}{c r_H^2}~.
\end{gather}
Here, $\Omega_H$ is the angular speed at the event horizon.
The form of the metric (\ref{abh}) looks similar to the
rotating BTZ black hole. However, we would like to note that the
lapse function $N(r)$ is different from that of the BTZ black hole
which is given by $N^2 = (r^2-r_+^2)(r^2-r_-^2)/(r^2 l^2)$.

We now consider a real scalar field $\varphi$ as a test field
in the background,
Eq.~(\ref{abh}), and investigate its behavior near the horizon.
First of all, the action of $\varphi$ is evaluated as
\begin{align}
S[\varphi]
&= - \int d^3 x \sqrt{-g} g^{\mu\nu} 
             \partial_\mu \varphi \partial_\nu \varphi 
\notag \\
&=  \int dt dr  \, r \!\! \int d \phi \, \varphi
\bigg( 
	- \frac{1}{N^2} ( \partial_t + \Omega_0 \partial_\phi)^2 
	+\frac{1}{r} \partial_r r N^2 \partial_r
	+\frac{1}{r^2} \partial_\phi^2
\bigg) \varphi ~.
\label{saction}
\end{align}
If we perform a wave decomposition of $\varphi$ as $\varphi = \sum_m
\varphi_m e^{im \phi}$, where $m$ is the angular quantum number
and $\varphi_m$ depends on the coordinates,
$t$ and $r$, then we see that the action is reduced to a
two-dimensional effective theory with an infinite collection of fields
labeled by $m$.  Next, in order to see what happens near the horizon,
it is helpful to take a transformation to the tortoise coordinate
$r^*$, which, in our case, is defined by
\begin{align}
\frac{\partial r^*}{\partial r} =
\frac{1}{N^2} \equiv \frac{1}{f(r)} ~, 
\end{align}
and leads to $\int dr = \int dr^* f(r(r^*))$.  In the region near the
horizon, $f(r(r^*))$ appears to be a suppression
factor vanishing exponentially fast, and thus the terms in the action
which do not have some factor compensating it can be ignored.  
In the present case, one can easily see that the last term
in the second line of Eq.~(\ref{saction}) is suppressed,
and hence the action near the horizon becomes
\begin{align}
S[\varphi] = \sum_{m \ge 0} \int dt dr \, r \,
  \varphi^*_m 
  \left( - \frac{1}{f} (\partial_t + i m \Omega_0)^2 
          + \partial_r f \partial_r 
  \right)
  \varphi_m ~,
\label{action}
\end{align}
where $\varphi^*_m \equiv \varphi_{-m}$.  Now it is not so difficult
to find that this near horizon action describes an infinite set of 
massless two-dimensional complex scalar fields in the following 
background:
\begin{gather}
ds^2 = - f(r) dt^2 + \frac{1}{f(r)} dr^2~,   \notag \\
\Phi = r ~, \quad A_t = -\Omega_0 ~, \quad A_r = 0~,
\label{2dbg}
\end{gather}
where $\Phi$ is the two-dimensional dilaton field.

Having the two-dimensional effective field theory near the horizon
(\ref{action}) and the two-dimensional background (\ref{2dbg}), we 
move on to the problem of Hawking radiation following
the approach based on the anomaly cancellation proposed in
\cite{rw,iso1}.  The anomaly approach of \cite{rw} begins with an
observation 
that, since the horizon is a null hypersurface, all ingoing
(left moving) modes at the horizon can not classically affect physics
outside the horizon.  This implies that they may be taken to be out of
concern at the classical level and thus the effective two-dimensional
theory becomes chiral, that is, the theory only of outgoing (right
moving) modes.  If we now perform the path integration of right moving
modes, the resulting quantum effective action becomes anomalous under
the gauge or the general coordinate transformation, due to the absence
of the left moving modes. However, such anomalous behaviors are in
contradiction to the fact that the underlying theory is not anomalous.
The reason for this is simply that we have ignored the quantum effects
of the classically irrelevant left moving modes at the horizon.  Thus
anomalies must be cancelled by including them.  

As stated above, the anomalies are localized at the
horizon $r_H$.  However, for actual computation, 
it is convenient to regard the quantum effective
action to be anomalous in an infinitesimal slab, $r_H \le r \le r_H +
\epsilon$, which is the region near the horizon.  (The limit $\epsilon
\rightarrow 0$ is taken at the end of the calculation.)  This leads to
a splitting of the region outside the horizon, $r_H \le r \le \infty$,
into two regions, $r_H \le r \le r_H + \epsilon$ and $r_H + \epsilon
\le r \le \infty$.  Then we will have the gauge and the gravitational 
anomaly near the horizon, $r_H \le r \le r_H + \epsilon$.  The reason
for the appearance of the gauge anomaly is that the 
background (\ref{2dbg}) contains the effective $U(1)$ gauge field due to 
the isometry along the angular direction and the near horizon action 
(\ref{action}) is that of charged fields. 

We first consider the gauge anomaly.  Let us denote the $U(1)$ gauge
current as $J_\mu$.
Since the region outside the horizon has been divided into two
regions, it is natural to write the gauge current as a sum
\begin{align}
J^{\mu} = J_{(o)}^{\mu} \Theta_+(r) 
          +  J_{(H)}^{\mu} H(r) ~,
\label{jsplit}         
\end{align}
where $\Theta_+(r) = \Theta(r-r_H-\epsilon)$ and $H(r)=1-\Theta_+(r)$.
Apart from the near horizon region, the current is conserved
\begin{align}
\partial_r J_{(o)}^{r} =0 ~.
\label{joeq}
\end{align}
On the other hand, the current near the horizon is anomalous and obeys
the anomalous equation
\begin{align}
\partial_r J_{(H)}^{r} = \frac{m^2}{4 \pi} \partial_r A_t ~,
\label{jheq}
\end{align}
which is the form of two-dimensional consistent gauge anomaly
\cite{bertlmann,bz}.  Since these two equations in each region are
first order differential ones, they can be easily integrated as
\begin{align}
J_{(o)}^{r} &= c_o, \notag \\
J_{(H)}^{r} &= c_H 
   + \frac{m^2}{4\pi} \left( A_t(r) -A_t(r_H) \right),
\label{jsol}
\end{align}
where $c_o$ and $c_H$ are integration constants.  The constant
$c_o$ is the electric charge flux which is of our concern.

Now, if $W$ is the quantum effective action of the theory without
including the ingoing (left moving) modes near the horizon, then its
variation under a gauge transformation with gauge parameter $\zeta$ is
given by
\begin{align} 
-\delta W 
& =\int d^2 x \sqrt{-g} \; \zeta \nabla_{\mu} J^{\mu}
\notag \\
&= \int d^2 x \; \zeta 
\left[
      \partial_r \left( \frac{m^2}{4\pi}A_t H \right)  
    + \delta(r-r_H - \epsilon) 
      \left( J_{(o)}^{r} - J_{(H)}^{r} 
             + \frac{m^2}{4 \pi}A_t \right)
\right] ~,
\label{gvar}
\end{align} 
where Eqs.~(\ref{jsplit}), (\ref{joeq}), and (\ref{jheq}) have been
used for obtaining the second line.  However, as mentioned before, the 
full quantum effective action of the underlying
theory must have gauge invariance.  The full effective action includes
the quantum effects of the ingoing modes near the horizon, whose gauge
variation gives a term canceling the first term of (\ref{gvar}).  For
the gauge invariance, the coefficient of the delta function in
Eq.~(\ref{gvar}) should also vanish, and hence, by using
Eq.~(\ref{jsol}), we get
\begin{align}
c_o = c_H - \frac{m^2}{4\pi} A_t(r_H) ~.
\end{align}
For determining the charge flux $c_o$, the value of the
current at the horizon, $c_H$, should be fixed.  This is done by
imposing a condition that the covariant current \cite{bz} given by
$\tilde{J}^r = J^r + \frac{m^2}{4\pi} A_t(r) H(r)$ vanishes
at the horizon, which, as noted in \cite{iso2}, assures the regularity
of physical quantities at the future horizon. Then, the electric
charge flux canceling gauge anomaly is determined as
\begin{align}
c_o =  - \frac{m^2}{2 \pi} A_t(r_H) 
= \frac{m^2}{2\pi} \Omega_H = \frac{m^2 B c}{2\pi A^2}~.
\label{co1}
\end{align}

The flux of the energy-momentum tensor radiated from the acoustic 
black hole is similarly determined through the cancellation of the 
gravitational anomaly.  First of all, like the splitting 
of Eq.~(\ref{jsplit}), we write the energy-momentum tensor as
\begin{align}
T^\mu_\nu = T^\mu_{\nu(o)}\Theta_+(r) +  T^\mu_{\nu(H)} H(r) ~.
\label{tsplit}         
\end{align}
Due to the presence of the gauge potentials and the dilaton in the
background (\ref{2dbg}), the energy-momentum tensor satisfies the
modified conservation equation \cite{iso1}.  What is of interest for
our problem is the conservation equation for the component $T^r_t$,
the energy-momentum flux in the radial direction.  
Apart from the near
horizon region, it is given by
\begin{align}
\partial_r T^r_{t(o)} = J^r_{(o)} \partial_r A_t ~.
\label{toeq}
\end{align}
In the near horizon region, we have
anomalous conservation equation \cite{iso1} as
\begin{align}
\partial_r T^r_{t(H)} 
=  J_{(H)}^r \partial_r A_t + A_t \partial_r J_{(H)}^r
 + \partial_r N^r_t ~,
\label{theq}
\end{align}
where $ N^r_t =( f^{\prime 2}+f f^{\prime\prime})/192\pi$.  (The prime
denotes the derivative with respect to $r$.)  The second term comes
from the gauge anomaly represented by the anomalous conservation
equation, while
the third term is due to the gravitational anomaly for the consistent
energy-momentum tensor \cite{aw}.  Now it is not a difficult task to
integrate Eqs.~(\ref{toeq}) and (\ref{theq}) and obtain
\begin{align}
T^r_{t(o)} &= a_o + c_o  A_t ~, \notag \\
T^r_{t{(H)}} &= a_H + \int^r_{r_H} dr \partial_r 
    \left( c_o A_t + \frac{m^2}{4\pi}A_t^2 + N^r_t \right) ~,
\label{tsol}
\end{align}
where $a_o$ and $a_H$ are integration constants. Here $a_o$ is the
energy flux which we are interested in.

The variation of quantum effective action $W$ under
a general coordinate transformation in the time direction with a
transformation parameter $\xi^t$ is obtained as
\begin{align}
- \delta W 
&= \int d^2x \sqrt{-g} \; \xi^t \nabla_\mu T^\mu_{t} 
\notag \\
&= \int d^2x \; \xi^t
  \bigg[ c_o \partial_r A_t +
        \partial_r 
        \left[ \left( \frac{m^2}{4\pi} A_t^2 + N^r_t \right) H
        \right]       
\notag \\
&+ \left( T^r_{t~(o)} - T^r_{t~(H)} 
         + \frac{m^2}{4\pi}A_t^2+N^r_t
   \right) \delta(r-r_H -\epsilon) 
     \bigg] ~.
\end{align}
The first term in the second line is purely the classical effect of
the background electric field for constant current flow.  The second
term is cancelled by including the quantum effect of the ingoing modes
as is the case of gauge anomaly.  The last term gives non-vanishing
contribution at the horizon and is also required to vanish for the
general covariance of the full quantum effective action.  This
requirement leads us to have the following relation.
\begin{align}
a_o = a_H +\frac{m^2}{4\pi}A_t^2(r_H)- N^r_t(r_H) ~,
\end{align}
where the solution Eq.~(\ref{tsol}) has been used.  For determining
$a_o$, we first need to know the value of $a_H$, which is fixed by
imposing a condition that the covariant energy-momentum tensor
vanishes at the horizon for regularity at the future horizon
\cite{iso2}.  Then, from the expression of the covariant
energy-momentum tensor \cite{bz,bk}, $\tilde{T}^r_t = T^r_t
+\frac{1}{192\pi} (f f'' -2(f')^2)$, the condition
$\tilde{T}^r_t(r_H)=0$ gives
\begin{align}
a_H= \frac{\kappa^2}{24 \pi} = 2N^r_t(r_H) ~,
\end{align}
where $\kappa$ is the surface gravity at the horizon,
\begin{align}
\kappa = \frac{1}{2} \partial_r f |_{r=r_H} =
\frac{1}{r_H}~.
\end{align}
Here, from the relation $T_H = \kappa / 2 \pi$, we see that
the Hawking temperature of the acoustic black hole is
\begin{align}
T_H = 
\frac{1}{2 \pi r_H}~,
\end{align}
which is the correct value.  Having the value of $a_H$, the
flux of the energy-momentum tensor is finally determined as
\begin{align}
a_o 
&= \frac{m^2}{4\pi}A_t^2(r_H) +N^r_t(r_H) \notag \\
&= \frac{m^2}{4\pi} \Omega_H^2 +\frac{\pi}{12} T_H^2 ~.
\label{tflux}
\end{align}

We have obtained the flux of electric charge,
Eq.~(\ref{co1}), and energy-momentum tensor,
Eq.~(\ref{tflux}), via the method of anomaly cancellation.  
If they are really the Hawking fluxes, they should coincide 
with the usual thermal fluxes of Hawking 
(black body) radiation from the black
hole.  Although the radiation in the case of bosons should be treated,
we simply consider the fermion case in order to avoid the
superradiance problem.  The Hawking distribution for fermions is given
by the Planck distribution at the Hawking temperature with 
electric chemical potential for the charge $m$ of the
fields radiated from the black hole,
\begin{align}
N_{m}(\omega) = 
\frac{1}{e^{(\omega - m \Omega_H)/T_H} +1} ~.
\end{align}
Then the electric charge flux of Hawking radiation, 
say $F_a$, and the energy-momentum flux of Hawking radiation, 
say $F_E$, can be obtained as
\begin{align}
F_a &= m \int^\infty_0 \frac{d\omega}{2\pi}
		( N_{m}(\omega) - N_{-m}(\omega) ) 
\notag \\
	&= \frac{m^2}{2\pi}\Omega_H~, \\
F_E &= \int^\infty_0 \frac{d\omega}{2\pi} \omega
		( N_{m}(\omega) + N_{-m}(\omega) ) 
\notag \\
	&= \frac{m^2}{4\pi} \Omega_H^2 +\frac{\pi}{12} T_H^2  ~.
\end{align}
We see that the thermal fluxes exactly match with (\ref{co1})
and (\ref{tflux}).
This implies that the fluxes of Hawking radiation from the black
hole are capable of canceling the gauge and
the gravitational anomalies at the horizon.

In summary, we have obtained the fluxes of Hawking radiation
from the three-dimensional acoustic black hole by using the method
of anomaly cancellation.  It has been shown that the resulting fluxes
are precisely the thermal fluxes from the two-dimensional black body
radiation at the Hawking temperature.  Therefore, our work confirms 
that the anomaly analysis proposed in \cite{rw,iso1} is valid for the
acoustic geometry modeling the draining bathtub.

\section*{Acknowledgments}
This work was supported by the Science Research Center Program of the
Korea Science and Engineering Foundation through the Center for
Quantum Spacetime (CQUeST) of Sogang University with grant number
R11-2005-021.  The work of H.S. was supported in part by grant
No.~R01-2004-000-10651-0 from the Basic Research Program of the Korea
Science and Engineering Foundation (KOSEF).


\begin{thebibliography}{10}
\bibitem{hawking}
  S.~W.~Hawking,
  ``Particle Creation By Black Holes,''
  Commun.\ Math.\ Phys.\  {\bf 43} (1975) 199
  [Erratum-ibid.\  {\bf 46} (1976) 206].
\bibitem{rw}
  S.~P.~Robinson and F.~Wilczek,
  ``A relationship between Hawking radiation and gravitational 
  anomalies,''
  Phys.\ Rev.\ Lett.\  {\bf 95} (2005) 011303
  [arXiv:gr-qc/0502074].
\bibitem{cf}
  S.~M.~Christensen and S.~A.~Fulling,
  ``Trace Anomalies And The Hawking Effect,''
  Phys.\ Rev.\  D {\bf 15} (1977) 2088.
\bibitem{iso1}
  S.~Iso, H.~Umetsu and F.~Wilczek,
  ``Hawking radiation from charged black holes via gauge and 
  gravitational anomalies,''
  Phys.\ Rev.\ Lett.\  {\bf 96} (2006) 151302
  [arXiv:hep-th/0602146].
\bibitem{iso2}
  S.~Iso, H.~Umetsu and F.~Wilczek,
  ``Anomalies, Hawking radiations and regularity in rotating black 
  holes,''
  Phys.\ Rev.\  D {\bf 74} (2006) 044017
  [arXiv:hep-th/0606018].
\bibitem{murata}
  K.~Murata and J.~Soda,
  ``Hawking radiation from rotating black holes and gravitational  
  anomalies,''
  Phys.\ Rev.\  D {\bf 74} (2006) 044018
  [arXiv:hep-th/0606069].
\bibitem{vd}
  E.~C.~Vagenas and S.~Das,
  ``Gravitational anomalies, Hawking radiation, and spherically 
  symmetric black holes,''
  JHEP {\bf 0610} (2006) 025
  [arXiv:hep-th/0606077].
\bibitem{setare}
  M.~R.~Setare,
  ``Gauge and gravitational anomalies and Hawking radiation of 
  rotating BTZ black holes,''
  Eur.\ Phys.\ J.\  C {\bf 49} (2007) 865
  [arXiv:hep-th/0608080];
  Q.~Q.~Jiang, S.~Q.~Wu and X.~Cai,
  ``Hawking radiation from the (2+1)-dimensional BTZ black holes,''
  arXiv:hep-th/0701048.
\bibitem{xu}
  Z.~Xu and B.~Chen,
  ``Hawking radiation from general Kerr-(anti)de Sitter black 
  holes,''
  Phys.\ Rev.\  D {\bf 75} (2007) 024041
  [arXiv:hep-th/0612261].
\bibitem{iso3}
  S.~Iso, T.~Morita and H.~Umetsu,
  ``Quantum anomalies at horizon and Hawking radiations in 
  Myers-Perry black holes,''
  arXiv:hep-th/0612286.
\bibitem{jiang1}
  Q.~Q.~Jiang and S.~Q.~Wu,
  ``Hawking radiation from rotating black holes in anti-de Sitter 
  spaces via gauge and gravitational anomalies,''
  Phys.\ Lett.\  B {\bf 647} (2007) 200
  [arXiv:hep-th/0701002].
\bibitem{jiang2}
  Q.~Q.~Jiang, S.~Q.~Wu and X.~Cai,
  ``Hawking radiation from the dilatonic black holes via 
  anomalies,''
  Phys.\ Rev.\  D {\bf 75} (2007) 064029
  [arXiv:hep-th/0701235].
\bibitem{kui}
  X.~Kui, W.~Liu and H.~Zhang,
  ``Anomalies of the Achucarro-Ortiz black hole,''
  Phys.\ Lett.\  B {\bf 647} (2007) 482
  [arXiv:hep-th/0702199].
\bibitem{sk}
  H.~Shin and W.~Kim,
  ``Hawking radiation from non-extremal D1-D5 black hole via anomalies,''
  arXiv:0705.0265 [hep-th].
\bibitem{Peng:2007pk}
  J.~J.~Peng and S.~Q.~Wu,
  ``Hawking radiation from the Schwarzschild black hole with a global
  monopole via gravitational anomaly,''
  arXiv:0705.1225 [hep-th].
\bibitem{Jiang:2007mi}
  Q.~Q.~Jiang,
  ``Hawking radiation from black holes in de Sitter spaces,''
  arXiv:0705.2068 [hep-th].
\bibitem{Chen:2007pp}
  B.~Chen and W.~He,
  ``Hawking Radiation of Black Rings from Anomalies,''
  arXiv:0705.2984 [gr-qc];
  U.~Miyamoto and K.~Murata,
  ``On Hawking radiation from black rings,''
  arXiv:0705.3150 [hep-th].
\bibitem{Jiang:2007pe}
  Q.~Q.~Jiang, S.~Q.~Wu and X.~Cai,
  ``Anomalies and de Sitter radiation from the generic black holes in de
  Sitter spaces,''
  arXiv:0705.3871 [hep-th].
\bibitem{wp}
  S.~Q.~Wu and J.~J.~Peng,
  ``Hawking radiation from the Reissner-Nordstr\'{o}m black hole with a
  global monopole via gravitational and gauge anomalies,''
  arXiv:0706.0983 [hep-th].
\bibitem{das}
  S.~Das, S.~P.~Robinson and E.~C.~Vagenas,
  ``Gravitational anomalies: a recipe for Hawking radiation,''
  arXiv:0705.2233 [hep-th].
\bibitem{iso4}
  S.~Iso, T.~Morita and H.~Umetsu,
  ``Higher-spin currents and thermal flux from Hawking radiation,''
  arXiv:hep-th/0701272;
  S.~Iso, T.~Morita and H.~Umetsu,
  ``Fluxes of Higher-spin Currents and Hawking Radiations from Charged Black
  Holes,''
  arXiv:0705.3494 [hep-th].
\bibitem{hms}
  G.~T.~Horowitz, J.~M.~Maldacena and A.~Strominger,
  ``Nonextremal Black Hole Microstates and U-duality,''
  Phys.\ Lett.\  B {\bf 383} (1996) 151
  [arXiv:hep-th/9603109].
\bibitem{visser}
  M.~Visser,
  ``Acoustic black holes: Horizons, ergospheres, and Hawking radiation,''
  Class.\ Quant.\ Grav.\  {\bf 15} (1998) 1767
  [arXiv:gr-qc/9712010].
\bibitem{unruh}
  W.~G.~Unruh,
  ``Experimental black hole evaporation,''
  Phys.\ Rev.\ Lett.\  {\bf 46} (1981) 1351.
\bibitem{kko}
  S.~W.~Kim, W.~T.~Kim and J.~J.~Oh,
  ``Decay Rate and Low Energy Near Horizon Dynamics of Acoustic Black Holes,''
  Phys.\ Lett.\  B {\bf 608} (2005) 10
  [arXiv:gr-qc/0409003];
  W.~T.~Kim, E.~J.~Son, M.~S.~Yoon and Y.~J.~Park,
  ``Statistical entropy and superradiance in 2+1 dimensional acoustic black
  holes,''
  J.\ Korean Phys.\ Soc.\  {\bf 49} (2006) 15
  [arXiv:gr-qc/0504127].
\bibitem{basak}
  S.~Basak and P.~Majumdar,
  ```Superresonance' from a rotating acoustic black hole,''
  Class.\ Quant.\ Grav.\  {\bf 20} (2003) 3907
  [arXiv:gr-qc/0203059];
  E.~Berti, V.~Cardoso and J.~P.~S.~Lemos,
  ``Quasinormal modes and classical wave propagation in analogue black
  holes,''
  Phys.\ Rev.\  D {\bf 70} (2004) 124006
  [arXiv:gr-qc/0408099];
  S.~Lepe and J.~Saavedra,
  ``Quasinormal modes, superradiance and area spectrum for 2+1 acoustic  
  black holes,''
  Phys.\ Lett.\  B {\bf 617} (2005) 174
  [arXiv:gr-qc/0410074].
\bibitem{bertlmann}
  R.~A.~Bertlmann,
  {\it Anomalies in Quantum Field Theory}
  (Oxford Science Publications, Oxford, 2000).
\bibitem{bz}
  W.~A.~Bardeen and B.~Zumino,
  ``Consistent And Covariant Anomalies In Gauge And Gravitational 
  Theories,''
  Nucl.\ Phys.\  B {\bf 244} (1984) 421.
\bibitem{aw}
  L.~Alvarez-Gaume and E.~Witten,
  ``Gravitational Anomalies,''
  Nucl.\ Phys.\  B {\bf 234} (1984) 269.
\bibitem{bk}
  R.~A.~Bertlmann and E.~Kohlprath,
  ``Two-dimensional gravitational anomalies, Schwinger terms and 
  dispersion relations,''
  Annals Phys.\  {\bf 288} (2001) 137
  [arXiv:hep-th/0011067].
\end{thebibliography}
\end{document}